\documentclass[showpacs, preprintnumbers, superscriptaddressprl, twocolumn, showpacs, floatfix, letterpaper, groupedaddress]{revtex4}
\usepackage{graphicx}
\usepackage{amsfonts}
\usepackage{amsmath, amsthm, amssymb}

\begin{document}
  
\newtoks\slashfraction
\slashfraction ={0.7}
\def\slash#1{\setbox0\hbox{$ #1 $}
\setbox0\hbox to \the\slashfraction \wd0{\hss \box0}\not\box0}

\def\dslash {\not{\hbox{\kern-2pt $\partial $}}}
\def\Dslash {\not{\hbox{\kern-4pt $D$}}}
\def\Qslash {\not{\hbox{\kern-3.5 pt $Q$}}}
\def\pslash {\not{\hbox{\kern-2.3 pt $ p$}}}
\def\kslash {\not{\hbox{\kern-2.3 pt $ k$}}}
\def\qslash {\not{\hbox{\kern-1.3 pt $ q$}}}
\def\Aslash {\not{\hbox{\kern-3.0 pt $ A$}}}

\newcommand{\bfk}{{\bf k}}  
\newcommand{\bfx}{{\bf x}} 
\newcommand{\bfp}{{\bf p}}

\title{Production of scalar particles in expanding spacetime}
\author{S. Ng}
\email{ng@stud.ntnu.no}
\affiliation{Department of Physics, Norwegian University of
Science and Technology, N-7491 Trondheim, Norway}
\date{Received    \today}


\begin{abstract}
In this paper, we investigate cosmological particle production using quantum field theory (QFT). We will consider how production of scalar particles can occur in an expanding universe. By introducing a time-dependent energy parameter representing the time evolution of the universe, the initial vacuum state will be excited. Consequently, creation of particles is present. Here, our focus is mainly creation of minimally coupled scalar particles in Minkowski spacetime.
\end{abstract}

\maketitle
\section{Introduction}\label{I}
Particle creation is an important topic in modern physics. The concept was given birth after the discovery of the inconsistency of vacuum. One of the first investigations of particle production by an expanding spacetime was first carried out by Bernard and Duncan in 1977. It showed that the concept of vacuum was incorrect, since a vacuum state does produce particles in a spacetime that undergoes some form of expansion. The creation of minimally coupled scalar particles, or free bosons, will be considered in this paper. There are many important reasons to consider scalar particles. These particles are certainly generated only by the geometry of the spacetime. In the calculation of the number of these peculiar bosons, we will see that these bosons mostly are zero-mode particles. Somehow, this may indicate the so-called spin-2 gravitons. The idea may also lead to an explanation of the vacuum energy divergences of bosons and fermions. Clearly, particle creation occurs in pairs, {\it i.e.} particles and anti-particles will be created as a pair. From the calculations in this paper, we will see that the initial vacuum state can be represented by a $2n$-excited final vacuum state.

\indent In Birrell \cite{Birrell}, this topic has been investigated in a two-dimensional Friedman-Robertson-Walker spacetime with a suitable line metric. In his analyse, the expansion of the spacetime is represented by the conformal scale factor. In this paper, however, we will investigate the same problem in term of QFT. In the derivation, produced particles will be identified as excitations of an initial vacuum state of a simple harmonic oscillator. Considering Klein-Gordon fields in an expanding spacetime represented by a time dependent energy term, minimally coupled scalar and neutral pairs of particles will be produced. Certainly, the evolution of the field is given by the so-called Bogoliubov transformation and the number of produced particles is represented by the absolute square of the Bogoliubov coefficient, $|\beta_\bfk|^2$.

\section{The model}\label{FT}
Here, we consider a simple model to see how excitations of a vacuum state can occur in a usual Minkowskian spacetime at distant past and distant future. Actually, since the term vacuum is not adequate, it should be replaced by ground state. But for simplicity,we will not be concerned by this, so in this model we may consider the ground state as a vacuum state. The ground state at $t\to -\infty$ is associated with an initial vacuum state which evolves in time. Indeed, the simplest way to start the investigation of this topic is a simple massless harmonic oscillator. For massless particles, there is equality between the momentum and the energy. Later, in consideration of massive bosons, the energy $\omega(t)$ can be written as a function of momentum $\bfk$ and mass $m$ in a quantum theoretical way. In our case, with the assumption of a time dependent energy the corresponding Hamiltonian acquires the same time dependence. By finding the probability of the transistion from the initial to the final vacuum state, excitations of the initial vacuum state will clearly be present. Obviously, this will also appear in calculation of the equation of the motion. Excitations of the initial vacuum state will appear in the calculations, which simply can represent production of particles due to time evolution. 

\indent Our starting point is a massless harmonic oscillator with a time dependent Hamiltonian given by 
\begin{equation}
\label{Hamilitonian}
	H(t) = \frac{1}{2}p(t)^2 + \frac{1}{2}\omega^2(t)q(t)^2,
\end{equation}
where $q(t)$ is the scalsr field and $p(t)$ is the momentum of the field. Both depend on the time $t$ and no other parameters in this simple model. The frequency is chosen to be 
\begin{equation}
\label{omegaAnsatz}
	\omega^2(t) = \omega^2_0 + \Delta \omega^2\, \tanh(\gamma t),
\end{equation}
which is our ansatz for the time evoulution of this model. This represents an expanding spacetime where far past and future are expressed as asymptotic regions $t\to -\infty$ and $t\to +\infty$. Since we are working with a massless field and the ansatz of the frequency is not a function of the spatial coordinate ${\bf x}$, spatial translation invariance is still valid and is a symmetry of this spacetime. It is thus convenient to consider the field as a quantum field, yielding
\begin{equation}
\label{eq:qft-form}
	q(t)= a \,u(t) + a^\dag u^*(t).
\end{equation}
Here, $a$ and $a^\dagger$ are the annihilation and creation operators and $u(t)$ is the time dependent function of the field. The Langrangian is $L = \dot q^2/2 - \omega^2(t) q^2/2$ and the canonical momentum is $p = {\partial L}/{\partial \dot q}= \dot q$. Here, we have set the mass parameter in the common harmonic oscillator equal to $1$, and $\gamma$ is a parameter adjusting the shape of the energy evolution. For the very early universe, {\it i.e.} $t\to -\infty$, we let $\omega(t) \to  \omega_1$, and for the distant future, $\omega(t) \to  \omega_2$, {\it i.e.} $t\to \infty$. To simplify analysis we only consider the asymptotic cases of $t\to \pm \infty$, corresponding to the case of {\it sudden limit}, {\it i.e.} $\gamma \to \infty$. Hence, the frequency will change its form at the point of $t=0$, {\it i.e.}
\begin{equation}
\label{eq:Omega-HarmOsc}
		\omega(t) = \begin{cases}
		\omega^2_0 - \Delta \omega^2  = \omega_1 & \text{for} \; t<0,  \\
		\omega^2_0 + \Delta \omega^2  = \omega_2 &  \text{for} \; t>0.
		\end{cases}
\end{equation}
From the Euler-Lagrange equations, we obtain the following equations
\begin{equation}
\label{bev-harm-osc}
	\ddot q(t)  + \omega(t)^2 q(t) = 0,
\end{equation}
where $\omega(t)$ is given by Eq. (\ref{eq:Omega-HarmOsc}). The derivative with respect to $t$ is denoted by an overdot. We consider $q(t)$ given by Eq. (\ref{eq:qft-form}) to be a time dependent quantum field on the form  
\begin{align*}
		q(t) &=  \frac{1}{\sqrt{2\omega_1}} (a_1e^{-i\omega_1 t} + a^\dagger_1e^{i\omega_1 t} ),  \hspace{0.1in} t<0 \\
		q(t) &=  \frac{1}{\sqrt{2\omega_2}} (a_2e^{-i\omega_2 t} + a^\dagger_2e^{i\omega_2 t} ),  \hspace{0.1in} t>0,
\end{align*}
where $a^\dagger$ and $a$ are the creation and annihilation operators and the prefactor comes from normalization. The field must obey the conditions of continuity $q(t= 0^-) = q(t= 0^+)$ and $\dot q(t= 0^-) = \dot q(t= 0^+)$. The conditions give the following equations,
\begin{equation}
\label{eq:aadag-lign1}
		\frac{ a_1 +a^\dagger_1}{\sqrt{2\omega_1}}  =	\frac{   a_2 +a^\dagger_2 }{\sqrt{2\omega_1}} 
\end{equation}
and 
\begin{equation}
\label{eq:aadag-lign2}
		\frac{\omega_1}{\sqrt{2\omega_1}} (a_1 -a^\dagger_1 ) =	\frac{\omega_2  }{\sqrt{2\omega_1}} ( a_2 -a^\dagger_2 )
\end{equation}
respectively. We define a new parameter $\eta$ such that
\begin{equation}
\label{eta}
	e^{\eta} = 	\sqrt {  \frac{\omega_2}{\omega_1} }.  	
\end{equation}
Writing Eqs. (\ref{eq:aadag-lign1}) and (\ref{eq:aadag-lign2}) in matrix form gives
\begin{equation}
\label{OmegaMatriserelasjon}
\left( {\begin{array}{*{20}c}
   {\frac{1}{{\sqrt {\omega _1 } }}} & {\frac{1}{{\sqrt {\omega _1 } }}}  \\
   { - \sqrt {\omega _1 } } & {\sqrt {\omega _1 } }  \\
\end{array}} \right)\left( {\begin{array}{*{20}c}
   {a_1 }  \\
   {a_1^\dag  }  \\
\end{array}} \right) = \left( {\begin{array}{*{20}c}
   {\frac{1}{{\sqrt {\omega _2 } }}} & {\frac{1}{{\sqrt {\omega _2 } }}}  \\
   { - \sqrt {\omega _2 } } & {\sqrt {\omega _2 } }  \\
\end{array}} \right)\left( {\begin{array}{*{20}c}
   {a_2 }  \\
   {a_2^\dag  }  \\
\end{array}} \right).
\end{equation}
By defining
\begin{equation*}
\left( {\begin{array}{*{20}c}
   {\frac{1}{{\sqrt {\omega _1 } }}} & {\frac{1}{{\sqrt {\omega _1 } }}}  \\
   { - \sqrt {\omega _1 } } & {\sqrt {\omega _1 } }   \\
\end{array}} \right)= {\bf \Omega}_1,   \hspace{0.2in} 
\left( {\begin{array}{*{20}c}
   {\frac{1}{{\sqrt {\omega _2 } }}} & {\frac{1}{{\sqrt {\omega _2 } }}}  \\
   { - \sqrt {\omega _2 } } & {\sqrt {\omega _2 } } \\
\end{array}} \right) = {\bf \Omega}_2.
\end{equation*}
the creation and annihilation operators for $t > 0$ from Eq. (\ref{OmegaMatriserelasjon}) can be writing as
\begin{align}
	\left( {\begin{array}{*{20}c}
   {a_2 }  \\
   {a_2^\dag  }  \\
	\end{array}} \right)
  	&= {\bf \Omega}_2^{-1}  {\bf \Omega}_1 
	\left( {\begin{array}{*{20}c}
   {a_1 }  \\
   {a_1^\dag  }  \\
	\end{array}} \right)  \notag \\ 
\label{aa-dagger-relasjon}
 	&=\frac{1}{2}\left( {\begin{array}{*{20}c}
   {\sqrt {\frac{{\omega _2 }}{{\omega _1 }}}  + \sqrt {\frac{{\omega _1 }}{{\omega _2 }}} } & {\sqrt {\frac{{\omega _2 }}{{\omega _1 }}}  - \sqrt {\frac{{\omega _1 }}{{\omega _2 }}} }  \\
   {\sqrt {\frac{{\omega _2 }}{{\omega _1 }}}  - \sqrt {\frac{{\omega _1 }}{{\omega _2 }}} } & {\sqrt {\frac{{\omega _2 }}{{\omega _1 }}}  + \sqrt {\frac{{\omega _1 }}{{\omega _2 }}} }  \\
\end{array}} \right)\left( {\begin{array}{*{20}c}
   {a_1 }  \\
   {a_1^\dag  }  \\
\end{array}} \right).
\end{align}
Eq. (\ref{aa-dagger-relasjon}) can be writting in a more compact form, reading 
\begin{equation}
\label{a2a2-dagger-relasjon}
\left( {\begin{array}{*{20}c}
   {a_2 }  \\
   {a_2^\dag  }  \\
\end{array}} \right) = \left( {\begin{array}{*{20}c}
   {\cosh \eta } & {\sinh \eta }  \\
   {\sinh \eta } & {\cosh \eta }  \\
\end{array}} \right)\left( {\begin{array}{*{20}c}
   {a_1 }  \\
   {a_1^\dag  }  \\
\end{array}} \right),
\end{equation}
where 
\begin{align*}
\cosh \eta  &= \frac{1}{2}\left(  \sqrt {\frac{{\omega _2 }}{{\omega _1 }}}  + \sqrt {\frac{{\omega _1 }}{{\omega _2 }}}     \right) \\ 
\sinh \eta  &= \frac{1}{2}\left(   \sqrt {\frac{{\omega _2 }}{{\omega _1 }}}  - \sqrt {\frac{{\omega _1 }}{{\omega _2 }}}     \right). 
\end{align*}
The inverse form of Eq. (\ref{aa-dagger-relasjon}) is 
\begin{equation}
\label{a1a1-dagger-relasjon}
\left( {\begin{array}{*{20}c}
   {a_1 }  \\
   {a_1^\dag  }  \\
\end{array}} \right) = \left( {\begin{array}{*{20}c}
   {\cosh \eta } & { - \sinh \eta }  \\
   { - \sinh \eta } & {\cosh \eta }  \\
\end{array}} \right)\left( {\begin{array}{*{20}c}
   {a_2 }  \\
   {a_2^\dag  }  \\
\end{array}} \right).
\end{equation}
Clearly, we have obtained an operator relation between the early universe and the present universe. The introduction of $\cosh \eta$ and $\sinh \eta$ above can be justified by associating with the relations 
\begin{align}
\label{assosisajon}
	x  \leftrightarrow a^\dagger \hspace{0.2in}{\rm and}  \hspace{0.2in} \frac{\partial}{\partial x}  \leftrightarrow a.
\end{align}
For an arbitary vacuum state $|\Omega  \rangle_ i$, the following relations are prevailing
\begin{align}
\label{eq:a1a2krav}
\begin{array}{*{20}c}
   {a _1  |\Omega  \rangle_ 1 =  0} , \\
   {a_2  |\Omega  \rangle_ 2 =  0}.  \\
\end{array}
\end{align}
Obviously, Eq. (\ref{a1a1-dagger-relasjon}) gives $\left(  \cosh \eta  a_2 - \sinh \eta a^\dag_2  \right)  |\Omega  \rangle_ 1  = 0$. Using the associations (\ref{assosisajon}), the annihilation  operator $a_2$ can be represented by $\partial / \partial x$ and the creation operator $a_2 ^\dag$ by $x$ in quantum mechanics. Fortunately, the initial vacuum state can be expressed as an excited state of $|\Omega\rangle$, reading
\begin{align}
	|\Omega \rangle_1    \sim  e^{ - \frac{1}{2}\tanh \eta \, x^2}	|\Omega \rangle_2.
\end{align}

\section{The vacuum states}
The conditions of the definition of a vacuum state for $t\to \pm \infty$ are given in Eq. (\ref{eq:a1a2krav}). From Eqs. (\ref{eq:a1a2krav}) and (\ref{a1a1-dagger-relasjon}), it is possible to find the relation between the vacuum states. From Eq. (\ref{a1a1-dagger-relasjon}), we have the relation 
\begin{equation}
\label{eq:cossinVac}
	(\cosh \eta \, a_2 - \sinh \eta \,a^\dagger_2)|\Omega\rangle _1 = a_1|\Omega\rangle _1 = 0.
\end{equation}
Obviously, $a_2|\Omega\rangle _1$ is non-zero. According to Eq. (\ref{eq:cossinVac}) we can make an ansatz of the initial vacuum state, reading
\begin{align}
\label{|Omega1}
	|\Omega\rangle _1 = \mathcal{N}e^{\frac{1}{2}\tanh \eta \, \, a_2^\dagger a_2^\dagger}|\Omega\rangle _2,
\end{align}
where $\mathcal{N}$ is the normalization constant. Using the association Eq. (\ref{assosisajon}), the normalization constant can easily be find. Utilizing series expansion of the exponential function in Eq. (\ref{|Omega1}), it can be written as a squeezed vaccum state with $\xi = -\tanh \eta$ in the form, 
\begin{align*}
|\xi \rangle  =  (1-|\xi^2|)^{1/4}   \sum\limits_{k = 0}^\infty  {\frac{(-\xi)^k \sqrt{(2k)!}}{2^k\,k!}} |2k\rangle, 
\end{align*}
where $|2k\rangle  = \frac{ (a^\dagger)^2 }{ \sqrt{(2k)!} }|\Omega\rangle $. Accordingly, the initial vacuum state (\ref{|Omega1}) reads
\begin{align*}
	|\Omega\rangle _1 &\approx \mathcal{N}     \sum \limits_{n = 0}^\infty  {\frac{1}{{n!}}} \left[ {\frac{1}{2}\tanh \eta } \right]^n \left( {a_2^\dag  } \right)^{2n} \left| \Omega  \right\rangle  _2  \\
&= \mathcal{N}    \sum\limits_{n = 0}^\infty  {\frac{{\sqrt {2n!} }}{{n!}}} \left[ {\frac{1}{2}\tanh \eta } \right]^n \frac{{\left( {a_2^\dag  } \right)^{2n} }}{{\sqrt {2n!} }}\left| \Omega  \right\rangle  _2,
\end{align*}
where we have used the coefficient of probability 
\begin{equation}
\label{eq:c2}
	c_{2n} = {\frac{{\sqrt {2n!} }}{{n!}}\left[ {\frac{1}{2}\tanh \eta } \right]^n }.
\end{equation}
Clearly, the initial vacuum state $|\Omega\rangle _1 $ is a kind of a $2n$ excitated state of $|\Omega \rangle_2$, since $a^\dag _2 a ^\dag _2 |\Omega \rangle_2 \sim |2n\rangle_2$. Utilizing the Stirling's formula 
\begin{equation}
\label{Stirling}
	x! \simeq \sqrt{2\pi x}x^x\, e^{-x}, \hspace{0.2in} x\to \infty
\end{equation}
can simplify the coefficient $c_{2n}$ in a more satisfying form $\sqrt {2n!}  = (4\pi n)^{1/4}2^n n^n e^{-n}$. Hence, $\sqrt {n!}  = \sqrt {2\pi n} \,n^n e^{-n}$. The prefactor in front of the $\tanh$-function in Eq. (\ref{eq:c2}) can be written in the form, $\frac{\sqrt {2n!} }{n!} \frac{1}{2^n} = \frac{  (4\pi)^{1/4} }{\sqrt {2\pi}n^{1/4}}$ for $n\to \infty$. We can use the condition of normalization,
\begin{equation}
\label{eq:NORM1}
{}_1\left\langle  {\Omega }
 \mathrel{\left | {\vphantom {\Omega  \Omega }}
 \right. \kern-\nulldelimiterspace}
 {\Omega } \right\rangle  _1    = 1,
\end{equation}
to rewrite Eq. (\ref{|Omega1}) into a geometric series. The rewritting is clear by utilizing the identity $\left( {\begin{array}{*{20}c}
   {2n}  \\
   n  \\
\end{array}} \right) =   \left( {\begin{array}{*{20}c}
   { - {1/2}}  \\
   n  \\
\end{array}} \right)( - 4)^n$. Consequently, 
\begin{align}
&\sum\limits_n {\left( {\begin{array}{*{20}c}
   {2n}  \\
   n  \\
\end{array}} \right)} \left( {\frac{{\tanh \eta }}{2}} \right)^{2n} \notag \\
&\,= \sum\limits_n {\left( {\begin{array}{*{20}c}
{ - {1/2}}  \\
   n  \\
\end{array}} \right)} \left( { - 2\tanh \eta } \right)^n  \notag\\
&\,= \left( {\frac{1}{{1 - \tanh ^2 \eta }}} \right)^\frac{1}{2}.
\end{align}
The normalization constant can be found by using Eq. (\ref{eq:NORM1}) and the relation $\mathcal{N} ^{-2} =  {1}/{\sqrt{1-\tanh ^2 \eta }}$. Hence, the normalization constant of the initial vacuum state reads $\mathcal{N} = {1}/{\sqrt{\cosh \eta}}$. The initial vacuum state can thus be expressed by the present vacuum state, yielding
\begin{equation*}
	|\Omega\rangle _1 = \frac{1}{\sqrt{\cosh \eta}}e^{\frac{1}{2}\tanh \eta \, \, a_2^\dagger a_2^\dagger}|\Omega\rangle _2.
\end{equation*}
From the definition of every excited state, $|n\rangle = \frac{1}{\sqrt{n!}}(a^\dagger)^n |\Omega \rangle$, the vacuum state $|\Omega\rangle _1$ is clearly a excited state of $|\Omega\rangle _2$,
\begin{equation}
\label{Vac12n}
	|\Omega\rangle _1 = \frac{1}{\sqrt{\cosh \eta}}  \sum\limits_{n=0}^\infty { \frac{\sqrt{2n!}}{n!} \left( \frac{1}{2} \tanh \eta \right)^n |2n\rangle_2},
\end{equation}
where $|2n\rangle _2 = \left(   (a_2^\dagger)^{2n}/  \sqrt{2n!}  \right)|\Omega\rangle _2$.

\indent According to Eq. (\ref{a2a2-dagger-relasjon}), we have the relation
\begin{equation}
\label{a-2Vakuum}
		(\cosh \eta \, a_1 + \sinh \eta a^\dagger_1)|\Omega\rangle _2 = a_2|\Omega\rangle _2 = 0.
\end{equation}
which implicitly gives the commutation rule reading 
\begin{equation}
\label{betingelseFa-2Vakuum}
	[a_2, e^{\frac{1}{2}\tanh\eta \, (a_2^\dagger)^2}] |\Omega \rangle_2 = e^{(\frac{1}{2}\tanh\eta)^n (a_2^\dagger)^{2}}.
\end{equation}
Obviously, Eq. (\ref{betingelseFa-2Vakuum}) is a consequence of Eq. (\ref{a-2Vakuum}), since this yields 
\begin{align*}
	[a_2, e^{\frac{1}{2}\tanh\eta \, (a_2^\dagger)^2}] &= \sum\limits_{n = 0}^\infty  { \frac{1}{n!}(\frac{1}{2}\tanh\eta)^n [a_2,(a_2^\dagger)^{2n}]}\\
		&= \sum\limits_{m = 0}^\infty {  \frac{1}{2}(\tanh\eta)(\frac{1}{2}\tanh\eta)^m }\\  
	&\,\,\,\,\times \frac{m+1}{(m+1)!} a_2^\dagger (a_2^\dagger)^{2n}  \\
		&= \tanh\eta \, a_2^\dagger  \sum\limits_{m = 0}^\infty { \frac{1}{m!}(\frac{1}{2}\tanh\eta)^m (a_2^\dagger)^{2m} }  ,
\end{align*}
by using the commutation rules $[A,BC] = [A,B]C + B[A,C] = 0$ and $[a_2,(a_2^\dagger)^M] = M\,(a_2^\dagger)^{M-1}$.

\section{The probability of excitations of the vacuum state}\label{Sannsynlighet}
It is interesting to find the probability of excitations in the initial vacuum state at $t\to -\infty$ due to the time evolution. In our case the excitations represent the number of produced particles in the harmonic oscillator, and the definition of the initial vacuum state, $|\Omega\rangle _1$, is the state at $t \to -\infty$. Vacuum state here means the lowest energy state of the Hamiltonian. The transition from $|\Omega\rangle _1$ to $|\Omega\rangle_2$ will initiate particle production depending on the energy relation, $\eta$ defined in Eq. (\ref{eta}). According to Eq. (\ref{Vac12n}), there is evident that $2n$ quanta will be created during the transition. Hence, the probability of the transition is found by placing the dual vector ${_2}\langle \Omega|$ into Eq. (\ref{Vac12n}). Due to the normalization (\ref{eq:NORM1}), the probability of the transition is found to be 
\begin{align}
	&P_{2n}(|\Omega\rangle _1  \to |2n\rangle _2 )  \notag \\
\label{P2n}
	&= | {_2\langle} 2n|\Omega\rangle _1|^2  =  \frac{1}{{\cosh \eta}} \left( {\begin{array}{*{20}c} {2n}\\
	n\\  \end{array}} \right)  \frac{1}{4^n}\left(\tanh \eta \right)^{2n},
\end{align}
where $n=0,1,2,..$ and $\left( {\begin{array}{*{20}c} {2n}\\ n\\  \end{array}} \right) = \frac{(2n)!}{(n!)^2}$ is a kind of a binomial coefficient, {\it i.e.} the number of ways of picking $n$ unordered outcomes from $2n$ possibilities. We see that $P_{2n}$ has a maximum value of ${\eta}$ for each $n$. If $\Delta \omega = 0$, {\it i.e.} $\eta = 0$, $P_0$ is  maximal at $\eta =0$. However, if we consider two particles are produced by the vacuum state $|\Omega\rangle _1$ when $t\to \infty$, $P_2$ is maximal at $\eta \simeq 1.146$, {\it i.e.} the present vacuum energy must be $1.318$ greater than the vacuum energy at the early universe.

\indent By considering $\eta$ and Eq. (\ref{P2n}), the vacuum state in the early universe, $|\Omega\rangle_1$, can be identified as an excited state of the present vacuum state, $|\Omega\rangle_2$.
\begin{figure}[h!]
	\centering{
	\resizebox{0.5\textwidth}{!}{
 	\includegraphics{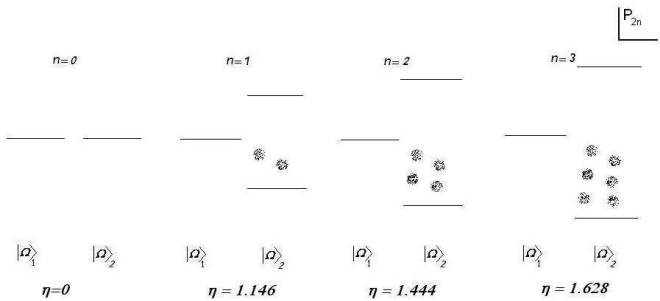}}
 	\caption{The $\eta$-value for $2n$ particles are produced by $|\Omega\rangle_1$ when $t\to \infty$. If there is no energy difference ($\eta =0$) between these two vacua, $|\Omega\rangle_1$ and $\Omega\rangle_1$, the Hamiltonian cannot be time dependent and therefore no particles can be produced. However, if $\eta>0$, there will produced particles in pairs.}
	\label{fig:P2antall}}
\end{figure}
\begin{figure}[h!]
	\centering{
	\resizebox{0.5\textwidth}{!}{
 	\includegraphics{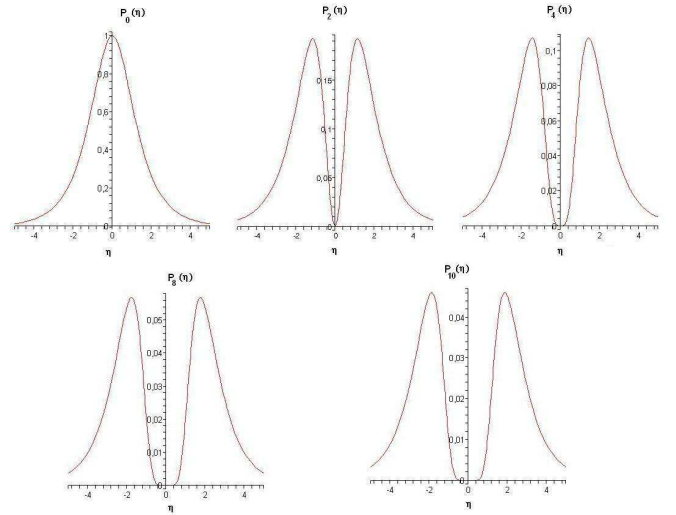}}
	\caption{The probability for zero, two, four, eight particles are produced by the time evolution of the state $|\Omega\rangle _1$, $t\to \infty$. When $|\eta|$ increases, the probability of more produced particles becomes greater.}
	\label{fig:PAntall}}
\end{figure}

\section{The general solution of the equation of motion}
\label{GenSolutionMasslessBos}
In order to find the expression of the field $q(t)$, we must solve the Euler-Langrange equation (\ref{bev-harm-osc}) 
\begin{equation}
\label{bevLignLoses}
		\ddot q(t) + (\omega_0^2 + \Delta^2 \tanh \gamma t)\,q(t) = 0		
\end{equation}
with
\begin{equation*}
	\mathop {\lim }\limits_{t \to  - \infty } q(t) = \frac{1}{\sqrt{2\omega_1}}(a_1 e^{-i\omega_1 t}+ a_1^\dagger e^{i\omega_1 t} ).	
\end{equation*}
Then, the question is how the field behaves when the time evolves to infinity, {\it i.e.} $q(t\to \infty)$. Let us generalize the field. We start with writing the field for $t\to -\infty$ in usual quantum field theoretical form, reading
\begin{equation}
\label{q(t)begy}
		q(t) = a_1 u_1(t)+  a_1^\dagger u^*_1(t),	
\end{equation}
where $u_1(t)$ is a time dependent orthonormal basis. We want to find $u_1(t)$ which satisfies the differential equation Eq. (\ref{bevLignLoses}). By inserting Eq. (\ref{q(t)begy}) into (\ref{bevLignLoses}), we certainly obtain 
\begin{equation}
\label{uBevLign}
		\ddot u_1(t) + (\omega_0^2 + \Delta \omega^2 \tanh\gamma t)u_1(t) = 0.		
\end{equation}
In order to identify Eq. (\ref{uBevLign}) with an equation with known solution, we introduce the new variable 
\begin{equation}
\label{tau}
	\tau(t) = \frac{1}{2}(1+\tanh(\gamma t)).
\end{equation}
Thus, the double derivative in time can be written in the form 
\begin{equation*}
	\frac{\mbox{d}^2}{\mbox{d}t^2} = 4\gamma^2\left\{  [\tau(1-\tau)]^2 \frac{\mbox{d}^2}{\mbox{d}\tau^2} +  \tau(1-\tau)(1-2\tau)] \frac{\mbox{d}}{\mbox{d}\tau} \right\}.
\end{equation*}
Writing Eq. (\ref{uBevLign}) with this new parameter, one certainly recognizes that this is a hypergeometric equation which reads
\begin{align}
	&4\gamma^2\left\{  [\tau(1-\tau)]^2 \frac{\mbox{d}^2}{\mbox{d}\tau^2} +  \tau(1-\tau)(1-2\tau)] \frac{\mbox{d}}{\mbox{d}\tau} \right\}u_1(\tau) \notag \\
\label{hypergeom} 
&+ \omega_0^2u_1 - \Delta \omega^2(1-2\tau)u_1(\tau)=0.
\end{align}
Clearly, $\tau$ goes to $0$ when $t\to -\infty$ and $\tau$ goes to $1$ when $t\to \infty$. The solution of Eq. (\ref{hypergeom}) has a typical form reading $u_1(\tau) = \tau^\nu \sum \limits_{n=0}^\infty {u_n \,  \tau^n}$, where $\nu$ is a parameter independent on $\tau$. The hypergeometric equation (\ref{hypergeom}) has three singular points $\tau = 0$, $\tau=1$ and $\tau = \infty$. We define the two time independent aymptotic vaules of $\omega(t)$, reading $\omega_1^2 = \omega_0^2 - \Delta\omega^2$ and $\omega_2^2 = \omega_0^2 + \Delta\omega^2$ for $t\to -\infty$ and $t\to \infty$ respectively. The index equation for $\tau=0$ by assuming $u_1 \sim \tau^{\nu_0}$ at the singularity $\tau=0$ reads
\begin{equation}
	\nu_0 (\nu_0 -1) + \nu_0 + \frac{ \omega_1^2}{4\gamma^2} =0.
\end{equation}
The solution acquires the form $u_1 \sim \tau^{\pm i\left(  \frac{ \omega_1}{2\gamma}\right)}$ for $\tau \to 0$. To obtain the standard form of QFT, we choose the negative sign of the indices of the solution. Hence, 
\begin{equation}
\label{tau-nu}
	u_1 \sim \tau^{-i\left(  \frac{ \omega_1}{2\gamma}\right)},
\end{equation}
For $\tau=1$, the solution of the field will go like $u_1 \sim (1-\tau)^{\nu_1}$. The index equation thus becomes 
\begin{equation}
	\nu_1 (\nu_1-1)+ \nu_1 + \left( \frac{ \omega_2^2}{4\gamma^2}\right) = 0.
\end{equation}
This gives us the solution of Eq. (\ref{hypergeom}) for the case of distant future, reading
\begin{equation}
\label{1-tau-nu}
	u_1 \sim (1-\tau)^{\pm i\left(  \frac{ \omega_2}{2\gamma}\right)}.
\end{equation}
Analog to (\ref{tau-nu}), we choose the right sign of the index of (\ref{1-tau-nu}) that satisfies the standard form of quantum field theory. Since $(1-\tau)^{\nu_1} \sim  {\rm exp}(-2\gamma t)$, it is therfore naturally to choose the positive sign of the index of (\ref{1-tau-nu}). Hence, the solution can be written in the form
\begin{equation}
\label{losningMf(tau)}
	u_1 = (1-\tau)^{ i\left(  \frac{ \omega_2}{2\gamma}\right)}  \tau^{ -i\left(  \frac{\omega_1}{2\gamma}\right)} f(\tau),
\end{equation}
where $f(\tau)$ is again a hypergoemetric function. $f(\tau)$ can be solved by inserting it back to the orginal hypergeometric differential equation (\ref{hypergeom}). Inserting (\ref{losningMf(tau)}) into Eq. (\ref{hypergeom}) certainly gives 
\begin{align}
	f'' + \frac{1}{\tau (1 - \tau )}  & \left[ \left( 1   - \frac{i\omega _2 }{\gamma } \right) \right.\notag\\
\label{hypergeom-m-f(tau)}
		&\left. - \left( 2 - \frac{i}{\gamma }(\omega _1  + \omega _2 )\tau  \right) \right]f' + \Upsilon f = 0,
\end{align}
where $'$ denotes the derivative with respect to $\tau$. $\Upsilon$ is a $\tau$-independent term coming from the last term in Eq. (\ref{hypergeom}) and by deriving of $f(\tau)$, yielding 
\begin{equation*}
	\Upsilon  = \left[ \frac{i(\omega _1  + \omega_2 )}{2\gamma } - \frac{(\omega _1  + \omega _2 )^2 }{4\gamma } \right].
\end{equation*}
Certainly, Eq. (\ref{hypergeom-m-f(tau)}) is a hypergeometric differential equation with the form 
\begin{equation}
\label{generellForm}
	\tau(1-\tau)f'' + \left[ c - (a+b+1)\tau \right]f' - ab\,f = 0.
\end{equation}
By comparing the general form of hypergeometric equation (\ref{generellForm}) with our equation (\ref{hypergeom-m-f(tau)}), we simply obtain 
\begin{align}
\label{a}
		a &= -\frac{i}{2\gamma}( \omega_1 + \omega_2) = -\nu_0 - \nu_1, \\
\label{b}
		b &= 1 - \frac{i( \omega_1 + \omega_2)}{2\gamma} = 1 + \nu_0 -\nu_1,\\
\label{c}
		c &= 1-\frac{i\omega_1}{\gamma} = 1+ \nu_0.
\end{align}
These expressions enter in the hypergeometric function $f (\tau)= {\rm const}\,F(a,b:c,\tau)$ \cite{Abram}. Clearly, the solution of the hypergeometric differential equation contains singularities in $\tau =0$ and $\tau =1$, yielding $u_1(\tau)  \sim  \tau ^({ \frac{-i\omega_1}{2\gamma}}) (1-\tau)^ ({ \frac{i \omega_2}{2\gamma}} ) \,F(a,b;c; \tau)$.

\indent For $\tau \to 0$, {\it i.e.} $t\to -\infty$, we obtain the same form we had to begin with Eq. (\ref{q(t)begy}) since $F(a,b;c; \tau)$ goes to $1$. The frequency mode becomes 
\begin{equation}
\label{u1tau0}
	u_1(t) \sim   \,  e^{ -i\omega_1  t},\hspace{0.2in} t\to -\infty
\end{equation}
where we have used 
\begin{equation}
\label{tauGaaSom}
	\tau   \sim  e^{ 2 \gamma t}. 
\end{equation}
Considering the singular point $\tau \to 1$, {\it i.e.} $t\to \infty$, the frequency mode $u_1(\tau)$ will obtain a time evolution given by a hypergeometric function, 
\begin{equation}
\label{u1tau1}
	u_1(t) \sim  \, e^{ -i\omega_2  t} \, F_{lim},
\end{equation}
where $F_{lim}$ is given by 
\begin{align}
	 F_{lim} &\equiv   \mathop {\lim }\limits_{x \to 1^+}  F(a,b;c;\, \tau)   \notag \\
   		 &= \mathop {\lim }\limits_{x \to 1^+}\, \left[ \,\frac{\Gamma ( c) \Gamma (c - a - b)}{\Gamma (c - a)\Gamma (c - b)}\, F(a,b;a + b -c - 1;\,  1-\tau )  \right. \notag   \\ 
	&+ \, \frac{\Gamma ( c)\Gamma (a + b - c)}{\Gamma (a)\Gamma (b)} (1 - \tau )^{c - a - b} \notag \\
\label{u-FF}
         &\left. \,\times F(c - a, c - b; c - a - b + 1; \,1-\tau )    \,\right].
\end{align}
This function becomes unity when $\tau \to 0$. The relations which enter in Eq. (\ref{u-FF}) are 
\begin{align}
c-a-b = i\omega_2/\gamma, &\hspace{0.1in} c-a= 1 + \frac{i(\omega_2-\omega_1)}{2\gamma}\notag   \\
\label{abcKombinasjoner}
		c-b = \frac{i(\omega_2-\omega_1)}{2\gamma}, &\hspace{0.1in} a + b - c = -\frac{i\omega_2}{\gamma}.
\end{align}
Hence, for the case of $t\to \infty$, the solution will go like 
\begin{equation}
\label{tauGaaSom1}
		(1-\tau)  \sim  e^{ -2 \gamma t},
\end{equation}
when we transform $\tau$ back to $t$. Here, the hypergeometric functions $F(a,b;a + b - c - 1; \,1-\tau )$ and $F( c - a, c - b; c - a - b + 1; \, 1-\tau )$ will go to 1. Thus we obtain two terms in the solution of $u_1$ for $t\to \infty$, containing time independent coefficients, $K_-$ and $K_+$, in each of them. Certainly, we get the forms of solution for those two cases $t\to \pm \infty$. By using Eqs. (\ref{u1tau0}) and (\ref{tauGaaSom1}), the frequency mode acquire the forms, $u_1 (t) = C \,e^{-i \omega_1 t }$ for $t\to-\infty$ and $u_1 (t) =  C_1 \,e^{i\omega_2 t }K_+\,+ \, C_2 \, e^{-i\omega_2 t}K_-$ for $t\to+\infty$. We have used the relations given in Eq. (\ref{abcKombinasjoner}) and defined the two time independent products of gamma functions as  
\begin{equation}
\label{eq:K+}
		K_+ = \frac{\Gamma ( c) \Gamma ( c - a - b)}{\Gamma (c - a)\Gamma (c - b)} 
\end{equation}
and 
\begin{equation}
\label{eq:K-}
		K_- = \frac{\Gamma ( c)\Gamma (a + b - c)}{\Gamma (a)\Gamma (b)}.
\end{equation}
The constants $C$, $C_1$ and $C_2$ in the frequency must have a form so that the field $q$ satisfies the communtation relations, $[a, a^\dagger] = 1$ and $[a, a] = [a^\dagger, a^\dagger] = 0$. However, the constants also must be in such way that the Lorentz invariance of the field is conserved in the continuous form; {\it e.g.} if the frequencies $\omega_1$ and $\omega_2$ are dependent on the wave vector ${\bf k}$, the integral of the field over ${\bf k}$ must be Lorentz invariant. Consequently, the constants thus obtain the forms $C = {1}/{\sqrt{2\omega_1}}$, $C_1 = C_2=  {1}/{\sqrt{2\omega_2}}$, which satisfy the hypergeometric equation (\ref{hypergeom}).

\indent Now, we will write down the asymptotic solutions of the field. For $t \to -\infty$, we have that $\tau \sim e^{2\gamma t}$. Thus the frequency mode has the form given by (\ref{u1tau0}) is obtained, since 
\begin{equation*}
			\tau ^{\frac{-i\,\omega_1}{2\gamma} } \sim e^{-i \omega_1 t}.
\end{equation*}
And for $t \to \infty$, $(1-\tau) \sim e^{-2\gamma t} $. It thus gives $(1-\tau )^{\frac{-i\,\omega_2}{2\gamma} } \sim e^{i \omega_2 t}$, which is consistent with (\ref{u1tau1}). Consequently, the solution of the field will obtain the usual form for $t \to-\infty$, {\it i.e.} $q(t) = a_1 u_1 + a^\dagger_1 u^*_1$ and a modified form for $t \to \infty$. Indeed, 
\begin{equation}
\label{solutionUhele}
		u_1(t) =  
\begin{cases}
		\frac{1}{\sqrt{2\omega_1}} \,    e^{-i\omega_1 t }, \hspace{0.1in}   t\to -\infty.    \\[1.0ex]
\begin{array}{*{20}c}
   {\frac{1}{\sqrt{2\omega_2}} \sqrt{ \frac{ \omega_2}{ \omega_1} } \, K_- \,  e^{-i\omega_2 t }}  \\
   {+  \frac{1}{ \sqrt{2 \omega_2} } \sqrt{ \frac{\omega_2}{\omega_1} }\, K_+ \,  e^{i\omega_2 t }}  \\
\end{array},           \hspace{0.1in}  t\to \infty.\\
		\end{cases}
\end{equation}
where $\sqrt{\frac{ \omega_2}{ \omega_1}}\, K_-$ and $\sqrt{\frac{ \omega_2}{ \omega_1}}\, K_+$ are the so-called Bogoliubov coefficients for the massless bosons, denoting by $\alpha$ and $\beta$ respectively. From (\ref{solutionUhele}) we see that the expression $u_1(t)$ obey the Bogoliubov transformation with the form \cite{Birrell}
\begin{equation}
\label{uTransf}
		u_1=\alpha\, u_2 +\beta\, u_2^*, \hspace{0.2in}   	u_1^* = \alpha^* u_2^* + \beta^* u_2,
\end{equation}
where $u_i= \frac{1}{\sqrt{2\omega_i}}e^{-i\omega_it}$, $i=1,2$. The annihilation operator will do the same with the form
\begin{equation}
\label{aTransf}
		a_1=\alpha^* \,a_2 -  \beta^* \,a_2^\dagger,\hspace{0.2in}   	a_1^\dagger = \alpha a_2^\dagger - \beta a_2.
\end{equation}

\indent In order to consider massive bosons, we include mass term in our ansatz of the evolution of the energy $\omega(t)$, yielding 
\begin{equation}
\label{omegakm}
	\omega_i (\bfk)  \to  \omega_i(\bfk) = \sqrt{{\bf k}^2 + m^2_i},  
\end{equation}
where $i= 1,2$. As for the massless bosons, this ansatz of the frequency is not a function of the spatial coordinate ${\bf x}$. Thus, spatial translation invariance is still a symmetry here. The solution (\ref{solutionUhele}) will be the same, but now $\omega(t)$ and the solution of the field $u_1(t)$ for the massless case include a $\bfk$-dependence, {\it i.e.} $\omega(\bfk, t)$ and $u(\bfk, t) = \left(1/\sqrt{2\omega_1}\right) \, {\rm exp} \left(-i\omega_1 t\right) {\rm exp} \left(i \bfk \bfx \right)$ .

\begin{figure}[h!]
	\centering{
	\resizebox{0.38\textwidth}{!}{
 	\includegraphics{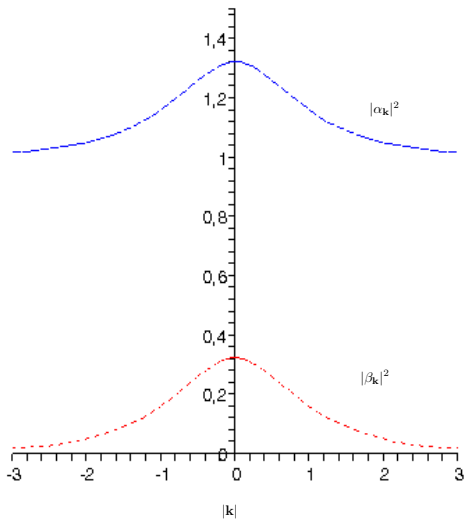}}
	\caption{The graphical representation of the Bogoliubov coefficients for the produced bosons. In this figure we compare the numerical coefficients with the expected analytical values, for a particular choice of parameters $m_1=1$, $\gamma = 10$ and $m_2=10$.} 
	\label{fig:bogoBoson}}
\end{figure}

\indent Thus, the Bogoliubov coefficients for the massive minimal coupled bosons read
\begin{equation}
\label{Bogo-koeff-alfa}
	\alpha = 	\sqrt{  \frac{\omega_2 }{\omega_1}  } \,     \frac{{\Gamma (1 - \frac{{i\,\omega _1 }}{\gamma })\Gamma ( - \frac{{i\, \omega _2 }}{\gamma })}}{{\Gamma ( -i\,\frac{{ \omega _1  +  \omega _2 }}{{2\gamma }})\Gamma (1 - i\,\frac{{\omega _1  +  \omega _2 }}{{2\gamma }})}} = \sqrt{\frac{ \omega_2}{\omega_1}}\, K_-
\end{equation}
\begin{equation}
\label{Bogo-koeff-beta}
	\beta = \sqrt{  \frac{\omega_2}{ \omega_1}  }    \,   \frac{{\Gamma (1 - \frac{{i\,\omega _1 }}{\gamma })\Gamma (  \frac{{i\,\omega _2 }}{\gamma })}}{{\Gamma (  i\, \frac{{ \omega _1  +  \omega _2 }}{{2\gamma }})\Gamma (1 + i\,\frac{{\omega _1  + \omega _2 }}{{2\gamma }})}}=  \sqrt{\frac{\omega_2}{\omega_1}}\, K_+.
\end{equation}
By using $\Gamma(x)\,\Gamma(1-x) = \frac{\pi}{\sin \pi x}$ \cite{Abram}, we thus obtain the moduli of the Bogoliubov coefficients
\begin{equation}
\label{alfa2}
	|\alpha|^2 = \frac{{\sinh ^2 (\pi \omega _ +  /\gamma )}}{{\sinh (\pi \omega _1 /\gamma )\sinh (\pi \omega _2 /\gamma )}},
\end{equation}
\begin{equation}
\label{beta2}
	|\beta|^2= \frac{{\sinh ^2 (\pi \omega _ -  /\gamma )}}{{\sinh (\pi \omega _1 /\gamma )\sinh (\pi \omega _2 /\gamma )}},	
\end{equation}
where we have defined the energy differences, $\omega_- =\frac{1}{2}(\omega_2-\omega_1)$ and $\omega_+ =\frac{1}{2}(\omega_2+\omega_1)$. Since the commutation relation of the operators for $t\to \infty$ is $[a_1,a_1^\dagger] = (|\alpha|^2 - |\beta|^2)[a_2,a_2^\dagger] = 1$, the normalization condition of the Bologoliubov coefficients reads
\begin{equation}
\label{abNorm}
	|\alpha|^2 - |\beta|^2 = 1.
\end{equation}
Grafically, the coefficients can be represented by Fig.\ref{fig:bogoBoson} as a function of momentum $|\bfk|$.
Certainly, the Bogoliubov coefficients we have found for a harmonic oscillator with the ansatz of the energy (\ref{omegakm}) are exactly the same results in Ref. \cite{Birrell}. Birrell (1982) has given a derivation in particle production by considering this topic as a cosmological problem describing by a metric for the Robert-Walker-universe with a conformal scale factor which corresponds the time evolution of our model of harmonic oscillator. 


\section{The properties of the Bogoliubov coefficients}
The Bogoliubov coefficients $\alpha_\bfk$ and $\beta_\bfk$ that we have found in Eqs. (\ref{Bogo-koeff-alfa}) and (\ref{Bogo-koeff-beta}) can be evaluated as
\begin{align}
	\alpha_{ij}   &=  (u_{1 i} , u_{2 j}) \notag\\
\label{ab}
	\beta_{ij}   &=   -(u_{1 i} , u_{2 j}^*),
\end{align}
where $x$ is the four-vector. The Bogoliubov transformations (\ref{uTransf}) and (\ref{aTransf}) are both diagonal and isotropic. Their coefficients prossess the following properties 
\begin{align}
	\sum \limits_{\bfk} { \alpha_{i\bfk} \alpha _{j \bfk}^*  -   \beta _{i\bfk}  \beta _{j\bfk}^* }  =\delta _{ij},\\
\sum \limits_{\bfk} { \alpha_{i\bfk} \beta _{j \bfk}  -   \beta_{i\bfk}  \alpha _{j\bfk} }  =0.
\end{align}
It follows that two spaces based on the two choices of modes $u_{1}$ and $u_{2}$ are different as long as $\beta_\bfk \neq 0$. As we have considered ealier, a vacuum state $|\Omega \rangle_1$ will not be annihilated by the annihilation operator $a_2$, {\it i.e.} $a_{2}| \Omega  {\rangle_1}  = \beta^*_\bfk |1 \rangle_1  \neq 0$. The state $|1 \rangle_1 $ is a excited state of the initial vacuum state $|\Omega \rangle_1$. Indeed, by using the Bogoliubov transformations the expectation value of the number operator $N_{2}= a_{2}^\dagger a_{2}$ of $u_{2}$-frequency mode particles in the initial state $|\Omega \rangle_1$ reads
\begin{equation}
\label{eq:exp-beta}
	{_1\langle} \Omega | N_{2}| \Omega  {\rangle_1}  ={|\beta|^2},
\end{equation}
which states that the initial vacuum state with frequency modes $u_{2}$ has been excited with $|\beta_\bfk|^2$ excitations. Simply stated, we can say that $|\beta|^2$ given in Eq. (\ref{beta2}) is the number of particles produced in the $u_{2}$-mode due to the time evolution of $\omega(t)$. It can thus interpret as production of minimally coupled scalar particles, {\it i.e.} bosons, in a time dependent harmonic oscillator. According to Eq. (\ref{eq:exp-beta}), the absolute square of $\beta_\bfk$ represents the expectation value of the number of particles present in the final state. From another point of view, the time evolution that we have assumed in our model can be considered in a pure cosmological context. Hence, the evolution represents the smoothly expanding motion of an aymptotically static universe. Indeed, the expansion excites the frequency modes of the (vacuum) field in that way we can identify excitations of the initial vacuum state $|\Omega_1 \rangle_1$ as created particles.


\section{Adiabatic expansion}\label{gamma0}
Considering an adiabatic expanding spacetime, the difference of the energies $\omega_-$ goes to zero. The number of produced particles represented by Eq. (\ref{beta2}) will thus vanish. Indeed, no particle can be created. This is a conformally trivial situation,{\it i.e.} a conformally invariant field propagating in a spacetime that is conformal to Minkowski spacetime. Hence, the particle creation occurs only when the conformal symmetry is broken by the presence of mass providing a length scale for the theory. The particle production can be identified the coupling of the spacetime expansion to the quantum field through the mass. The expansion regarded as a changing gravitational field will be an energy supplier to the perturbed scalar field modes. When we consider creation of field quanta with the changing gravitational field, it is natural to describe the particles as being produced in the asympotically static regions. Certainly, particle production can not occur in the asymptotically static regions. However, if $|\beta|^2$ represents a final density of particles which were not present in the state $|\Omega\rangle_1$, we would except that a measurment taking place at an intermediate time during the expansion era, would unveil a particle density in the interval from zero to the number of particles is produced given by the absolute square of $\beta_\bfk$. 

\indent This do not live up to scrutiny, however, if the spacetime is curved, there will no good definition of particles is generally available. Because of the special symmestry of the FRW spacetime, we can identify comoving observers who see the universe isotropically expands. In this case, it is reasonable to consider produced particles in the expansion region as the {\it excitation} of comoving particle observer. 

\indent Let us get back on track and arrive at the mathematical description. The adiabatic case certainly gives limits of Eqs. (\ref{eq:K+}) and (\ref{eq:K-}) regarded to the Bogoliubov coefficients (\ref{Bogo-koeff-alfa}) and (\ref{Bogo-koeff-beta}), {\it i.e.}
\begin{align}
\label{eq:K+K-1}
	K_+ &\to  \frac{\Gamma (1-\frac{i\omega}{\gamma}) \Gamma (\frac{i \omega}{\gamma})}{\Gamma(1)}   = \frac{-i( \omega_1- \omega_2)}{2\gamma} =  0 \\
\label{eq:K+K-2}
	K_- &\to  \frac{\Gamma (1-\frac{i \omega}{\gamma}) \Gamma (-\frac{i\omega}{\gamma})}{\Gamma (-\frac{i \omega}{\gamma})\Gamma (1-\frac{i\omega}{\gamma})}  = 1,
\end{align}
since $\omega_2= \omega_1 = \omega$ in ths case. In an adiabatic ecpanding spacetime, it is appropriate to assume no energy change, {\it. i.e.} $\omega_1=\omega_2$. The field $q(t)$ is approximately equal for both asymptotical cases. According to Eqs. (\ref{eq:K+K-1}) and (\ref{eq:K+K-2}), the solution (\ref{solutionUhele}) acquires the form 
\begin{align}
	u_1(t)= \frac{1}{\sqrt{2\omega_1}} e^{-i\omega_1 t}  = \frac{1}{\sqrt{2\omega_2}} e^{-i\omega_2 t}.
\end{align}
This clearly states no change in of the field. The field evolution satifes the normalization condition (\ref{abNorm}) according to Eqs. (\ref{eq:K+K-1}) and (\ref{eq:K+K-2}), yielding $1^2-0^2=1$. The number of created particles, {\it. i.e.} number of excitations, is represented by $|\beta|^2$. In an adiabatic expansion, the creation rate approachs to zero when the expansion rate go to zero. Here, it is appropriate to consider $\gamma$ as the expansion rate since it is the parameter of the time evolution. When $\gamma \to 0$, Eq. (\ref{beta2}) goes thus like
\begin{equation}
	[ \pi(\omega_2 - \omega_1 ) - \pi(\omega_2+\omega_1)]/\gamma = e^{-2\pi \omega_1/\gamma} \to 0,
\end{equation}
which states that the number of created quasi-particles rapidly approaches zero in an adiabatically expanding spacetime. 

\section{Sudden limit}
\subsection{Sudden limit}
For sudden limit the expansion rate increases dramatically in the time evolution, {\it i.e.} $\gamma \to \infty$. Certainly, the time parameter $\tau(t)$ given by (\ref{tau}) approaches to a step function. The terms on the hypergeometric function, $K_+$ and $K_-$, will go like 
\begin{align}
\label{eq:K+K-S1}
	K_+ &\sim  \frac{\gamma}{i \omega_2} \frac{-i (\omega_1 - \omega_2)}{2\gamma} = \frac{1}{2}(1-\frac{ \omega_1}{\omega_2}) \\
\label{eq:K+K-S2}
	K_- &\sim  \frac{\gamma}{-i \omega_2} \frac{-i ( \omega_1 -  \omega_2)}{2\gamma} = \frac{1}{2}(1+\frac{ \omega_1}{ \omega_2}).
\end{align}
In the sudden limit approximation, things become more dramatically. The initial vacuum state will accordingly be excited (or de-excited) due to the enormous change of the spacetime. The initial state energy $\omega_1$ evolves to $\omega_2$ for $t\to \infty$. According to the Eqs. (\ref{eq:K+K-S1}) and (\ref{eq:K+K-S2}), the solution (\ref{solutionUhele}) acquires the form 
\begin{align}
	u_1(t) =   \left\{   \begin{array}{*{20}c}
	   {\frac{1}{\sqrt{2\omega_1}  } e^{-i\omega_1 t},\,        \hspace{0.6in}   t\to -\infty}  \\[1.6ex]
	   {   \begin{array}{*{20}c}
   {\frac{1}{\sqrt{2\omega_2}}  \sqrt{\frac{\omega_2}{\omega_1}}  \left[ \frac{1}{2}(1+\frac{ \omega_1}{\omega_2})  e^{-i\omega_2 t}  \right.       }  \\
   { \,\left. +  \frac{1}{2}(1-\frac{ \omega_1}{\omega_2}) e^{-i\omega_2 t}  \right], \, \hspace{0.1in}  t \to \infty}  \\
\end{array}      } 
\end{array}  \right.
\end{align}
Fortunately, the field evolution satifes the normalization condition (\ref{abNorm}), reading 
\begin{align}
\label{eq:sjekkNormBet}
	\left| {\frac{1}{2}\sqrt {\frac{{\omega _2 }}{{\omega _1 }}} \left( {1 + \frac{{\omega _1 }}{{\omega _2 }}} \right)} \right|^2  - \left| {\frac{1}{2}\sqrt {\frac{{\omega _2 }}{{\omega _1 }}} \left( {1 - \frac{{\omega _1 }}{{\omega _2 }}} \right)} \right|^2  
	= \frac{{\frac{{\omega _1 }}{{\omega _2 }}}}{{\left| {\frac{{\omega _1 }}{{\omega _2 }}} \right|}},
\end{align}
where we have assumed that the energies for both asymptotic cases are positive. Since the field expansion changes in the case, the number of created particles can not be neglected. For sudden limit, the expansion is able to generate particle production. However, the bosons which are been created due to a sudden expansion of the spacetime is nevertheless very small. According to $|\beta_\bfk|^2$, there will be produced mostly zero-mode particles, {\it i.e.} $\bfk =0$. Therefore, in any calculation of the bosonic vacuum energy, which is a sum over $\bfk$, the created bosons will thus be invisibles in the final ({\it i.e.} $t\to \infty$) vacuum energy calculation, since the number of created bosons is mainly zero-mode particles.

\section{Concluding remarks}
In this papir, we have given a derivation of particle production in an expanding spacetime by using a simple harmonic oscillator model and the Bogoliubov formalism. We have shown that particle creation in a time-depending harmonic oscillator can occur in a spacetime with a suitable ansatz of the time evolution, corresponding to the motion of the expansion of the universe. Since the very beginning of the universe underwent an enormous expansion, this spontaneous production of particles played a vigorous part at that time. According to the calculation, created particles are equivalent to excitations of the initial vacuum state. In this model, we have only considered bosonic case of the cosmological particle production. 

\section{Acknowledgments}  
This article has been the first part of a diploma thesis at NTNU, Norwegian University of Science and Technology. The author want to express his gratitude to Professor K. Olaussen for helpful discussions along the way. This work was funded in part by the State Educational Fund and the Department of Physics, NTNU, in 2005. \\


\section{References}
\begin{thebibliography}{99}
\bibitem{Abram} M. Abramowitz and I. A. Stegun, {\it Handbook of mathematical functions with Formula, Graphs, and Mathematical Tables} (U.S. Department of Commerce, National Bureau of Stanards, 1964).
\bibitem{Nuno} N. D. Antunes, arXiv:hep-ph/0311031 (2003).
\bibitem{Birrell} N. D. Birrell and P. C. W. Davies, {\it Quantum Fields in Curved Space}, (Cambridge University Press, Cambridge, 1982).
\bibitem{Davies} P. C. W. Davies, J. Opt. B: Quantum Semiclass. Opt. {\bf 7} (2005).
\bibitem{Gasperini} M. Gasperini and  M. Giovannini, arXiv:hep-th/9502112 (1995).
\bibitem{Henry} R. W. Henry and S. C. Glotzer, {\it A squeezed-state primer}, Am. J. Phys. {\bf 56} 138 (1988).
\bibitem{Hong} J. Hong and A. Vilenkin, arXiv:gr-qc/0210034 (2002).\bibitem{Mandl} F. Mandl, G. Shaw, {\it Quantum Field Theory} (J. Wiley \& Sons, Inc., 1984).
\bibitem{koks} D. Koks, Phys. Rev. D {\bf 56}, 8 (1997).





\end{thebibliography}
\end{document}